\documentclass[pra,superscriptaddress,onecolumn,amsmath,amssymb]{revtex4}
\usepackage{amsfonts}
\usepackage{amssymb}
\usepackage{mathtools}
\usepackage{graphicx}
\usepackage[english]{babel}
\usepackage{subfigure}
\usepackage{mathrsfs}
\usepackage{color}
\usepackage[normalem]{ulem}
\usepackage{natbib}
\usepackage{bbold}
\usepackage{placeins}
\usepackage{braket}
\usepackage{soul,xcolor}
\usepackage{epstopdf}
\usepackage{tabu}
\usepackage{array}
\usepackage{bm}
\usepackage{upgreek}
\usepackage{multirow}
\usepackage[utf8]{inputenc}
\usepackage[colorlinks = truelinkcolor = blue, urlcolor  = blue, citecolor = blue, anchorcolor = blue]{hyperref}
\hypersetup{
	colorlinks   = true, 
	urlcolor     = black, 
	linkcolor    = blue, 
	citecolor   = blue 
}

\newcommand{\norm}[1]{\left\lVert#1\right\rVert}
\makeatletter

\begin{document}

\title{Distinguishing environment-induced non-Markovianity from subsystem dynamics}

\author{Subhashish Banerjee}
\email{subhashish@iitj.ac.in}
\affiliation{Indian Institute of Technology Jodhpur, Jodhpur 342011, India}

\author{Javid Naikoo}
\email{naikoo.1@iitj.ac.in}
\affiliation{Indian Institute of Technology Jodhpur, Jodhpur 342011, India}

\author{ R. Srikanth}
\email{srik@poornaprajna.org}
\affiliation{Poornaprajna Institute of Scientific Research, Bangalore - 562164, India}

\begin{abstract}
	\noindent
Quantum non-Markovianity modifies the environmental decoherence of a system.
This situation is enriched in complex  systems  owing to  interactions among  subsystems.  We consider the problem of distinguishing the multiple sources of non-Markovianity using a  simple  power spectrum  technique,  applied to a qubit interacting with another qubit via a Jaynes-Cummings type Hamiltonian and simultaneously subjected
to  some well known noise channels, such as, the random telegraph noise  and non-Markovian amplitude damping, which exhibit  both Markovian as well as  non-Markovian dynamics under different parameter ranges.	
\end{abstract}
\maketitle

\section{Introduction}  

Any practical implementation of quantum information processing demands
taking into account the effects  of the ambient environment, resulting
in     the     phenomena     of    decoherence     and     dissipation
\cite{breuer2002theory,sbbook,srikanth2008,OmkarSingleQubit,chandrashekar2007symmetries, BRC+08}.   Until
recently, noise  was predominantly  studied in  the \textit{Markovian}
(memoryless)  regime,  where  the  environmental time  scale  is  much
smaller than the  system time scale \cite{RHP14,VA17},
entailing the two  weak conditions: (a) the monotonic  fall of the
distinguishability $D(\mathcal{E}[\rho_1], \mathcal{E}[\rho_2])$ of two states
$\rho_j$,  where  $\mathcal{E}$  is   the  noise  superoperator;  (b)  the
CP-divisibility of  the noise dynamics, meaning  that the intermediate
map remains  completely positive,  essentially because the  system and
bath remain in an approximately product state during the evolution. It
is  known that  (a) implies  (b).  Recently,  \cite{bognaDisVsBF} have
argued  that  these  two   conditions  are  equivalent  for  bijective
maps.

With  a breakdown  of  condition  (a), there  is  an  increase in  the
distinguishability $D$,  causing a  recurrence or  ``backflow'' of
information back  from the  environment into  the system.   With a
breakdown  of condition  (b), the  intermediate map  is non-CP  (NCP),
essentially because the  system-bath interaction generates system-bath
entanglement. With the advancement of  technology, one is now able
to experimentally  go beyond  Markovian phenomena  and enter  into the
non-Markovian regime \cite{RHP10,RHP14,NMCrypto,NMDephasing1,NMDephasing2,javid,QuThermo,PLNOUN}.

A complex  dynamics, in general, will have a number of evolution patterns, in the form of different frequency components from different sources, superimposed to produce the resulting dynamics. This was seen in earlier works \cite{pradeep1,pradeep2} in the context of discrete time quantum walk (DTQW). There, the patterns superimposed on the position dynamics of the walker were the evolution due to its interaction degree of freedom traced out and noise due to an external source. Here, we elucidate the idea of disambiguating different sources of noise using a different model, viz., a qubit system evolved by a Jaynes Cummings type of evolution \cite{scully-zubairy,gsa} and a   noise channel, such as the random telegraph noise (RTN) channel \cite{daffner,vankampen}, non-Markovian  amplitude damping (NMAD) channel, and a concatenation of   these channels.   This could be very pertinent in complex  as well as engineered systems  where this scenario could be envisaged due to interaction among  subsystems.   Keeping this in mind, we consider a composite dynamics different from that studied in \cite{pradeep1}, in order to better illustrate the frequency-based noise disambiguation technique. The idea here is to associate the spectral power associated with different frequency components as a measure of  their relative strengths of non-Markovianity (CP-indivisibility).

\section{Non-Markovian Noise}
\textit{Random Telegraph Noise }(RTN):~
We briefly describe the features of RTN   noise, needed for our purpose. The random variable describing the noise fluctuation will be denoted by $\Omega(t)$ and by $M$ the mean.  The  autocorrelation   functions  and  the corresponding    Kraus    operators $K_i$    are   as follows:
\begin{eqnarray}
	\label{noiseprop}
	M[\Omega(t),\Omega(s)] = a^2 e^{-|t-s|/ \tau}, \qquad
	K_1    =    \sqrt{\frac{1+\Lambda(t)}{2}} I, \qquad
		K_2    =	\sqrt{\frac{1-\Lambda(t)}{2}} \sigma_3,
	\label{eq:rtn}
\end{eqnarray}
where 
\small
\begin{equation}\label{eq:lambda}
\Lambda(t) =	e^{-\gamma t}\left[ \textmd{cos}\left(\left[\sqrt{(\frac{2a}{\gamma})^2-1}\right]\gamma t \right)+
	\frac{\textmd{sin}\left(\left[\sqrt{(\frac{2a}{\gamma})^2-1}\right]\gamma t\right)}{\sqrt{(\frac{2a}{\gamma})^2-1}}\right],
\end{equation} 
\normalsize
represents  the  damped harmonic  function,  which
encodes both the Markovian and non-Markovian behaviour of the noise. 
The reduced dynamics of a qubit subjected to RTN can be described by  map $\mathcal{R} : \rho (t) = \mathcal{R}(t)[\rho(0)] = \sum_i K_i(t) \rho(0) K_i^\dagger(t)$, with the Kraus operators as given in 
Eq. (\ref{eq:rtn}).

In RTN, the  function $\Lambda(t)$  has  two regimes; the purely
damping  regime,  where  $a/\gamma<0.5$,  and  damped  oscillations  for
$a/\gamma>0.5$. Corresponding to these regimes of $\Lambda(t)$, one
observes   Markovian   and    non-Markovian   behavior,   respectively. \bigskip

 \textit{Non-Markovian Amplitude Damping} (NMAD):~ This model describes the dissipative interaction between a qubit and zero temperature Bosonic reservoir, which could be envisaged to be composed of a large number of harmonic oscillators. The channel  $\mathcal{A}_t$ acts  on a general input state $\rho = (\rho_{11}, \rho_{12}; \rho_{21}, \rho_{22})$ as:
 
            \begin{equation}\label{eq:mapNMAD}
               \mathcal{A}_t (\rho) = \begin{pmatrix}
                                                     1 - |G(t)|^2) \rho_{22}   &    G(t) \rho_{12} \\
                                                       G^*(t) \rho_{12}         &      |G(t)|^2) \rho_{22}
                                                \end{pmatrix}.
            \end{equation}
            The dynamics can be described by the following Kraus operators:
            \begin{equation}\label{eq:KrausA1A2}
            A_1 = \begin{pmatrix}
                             1 & 0 \\
                             0  &  G(t)
            \end{pmatrix}, \qquad  A_2 = \begin{pmatrix}
                                                          0   & \sqrt{1 - |G(t)|^2}\\
                                                          0  &  0
                                                       \end{pmatrix}.
            \end{equation}
           The reservoir spectral density can be assumed to be a Lorentzian $J(\omega) = \frac{\gamma_M \lambda^2}{2 \pi (\omega - \omega_c)^2 + \lambda^2}$, with width $\lambda$, peak frequency $\omega_c$ and $\gamma_M$ an  effective coupling constant.   The function $G(t)$ becomes 
            \begin{equation}\label{eq:Gt}
            G(t) = e^{-(\lambda - i \delta)t/2} \Big[\cosh(\Omega t/2) + \frac{\lambda - i \delta}{\Omega} \sinh(\Omega t/2)   \Big].
            \end{equation}
            Here, $\Omega = \sqrt{\lambda^2 - 2 i \delta \lambda - 4w^2 }$, $w = \gamma_M \lambda/2$, $\delta = \omega_0 - \omega_c$, and   $\omega_0$ is the natural frequency corresponding to the input qubit state.  The  condition $\lambda/\gamma_M \gg 1$ corresponds to the flat and weakly coupled spectrum and hence pertains to the Markovian dynamics. In other words, the dynamics is non-Markovian for $\lambda / \gamma_M \ll 1$.

Under a Markovian evolution $\mathcal{E}$,  given two
distinct states $\rho_1$ and  $\rho_2$, distance measures $\mathfrak{D}$
(such    as   relative    entropy   or    trace   distance)    satisfy
$\mathfrak{D}[\mathcal{E}(\rho),  \mathcal{E}(\sigma)] \le  \mathfrak{D}[\rho,
\sigma]$,  while   correlation  measures  $\mathfrak{C}$   (such  as
fidelity  or   mutual  information   with  a  third   system)  satisfy
$\mathfrak{C}[\mathcal{E}(\rho),            \mathcal{E}(\sigma)]           \ge
\mathfrak{C}[\rho,\sigma]$.  By  contrast, non-Markovian  dynamics can
violate  the above  monotonicity  property,  resulting in  information
backflow  from  the  environment   into  the  system,  manifesting  as
recurrence.  In this work, we  will use trace-distance (TD) applied to
a simple model as  the indicator  of distinguishability  of two  states.
Non-Markovian  backflow  has been  also  been  studied using  fidelity
\cite{rajagopal2010kraus},  relative entropy  \cite{DRS11} and  mutual
information.

\section{Non-Markovian witnesses: CP divisibility vs. distinguishablity}

As noted, non-Markovianity may be  indicated by observing the backflow
in  the   evolution  \cite{breuer2009measure,breuer2016colloquium}  or
departure from  CP-divisibility of the intermediate  map connecting the
density  operators  $\rho(t_2)$ and  $\rho(t_1)$  at  times $t_2$  and
$t_1$, with  $t_2>t_1$ \cite{RHP10,RHP14}. Both methods  are discussed
here and found to be equivalent for the noise models considered.
\subsection{CP-divisibility criterion}
\begin{figure}
	\centering
	\includegraphics[width=70mm]{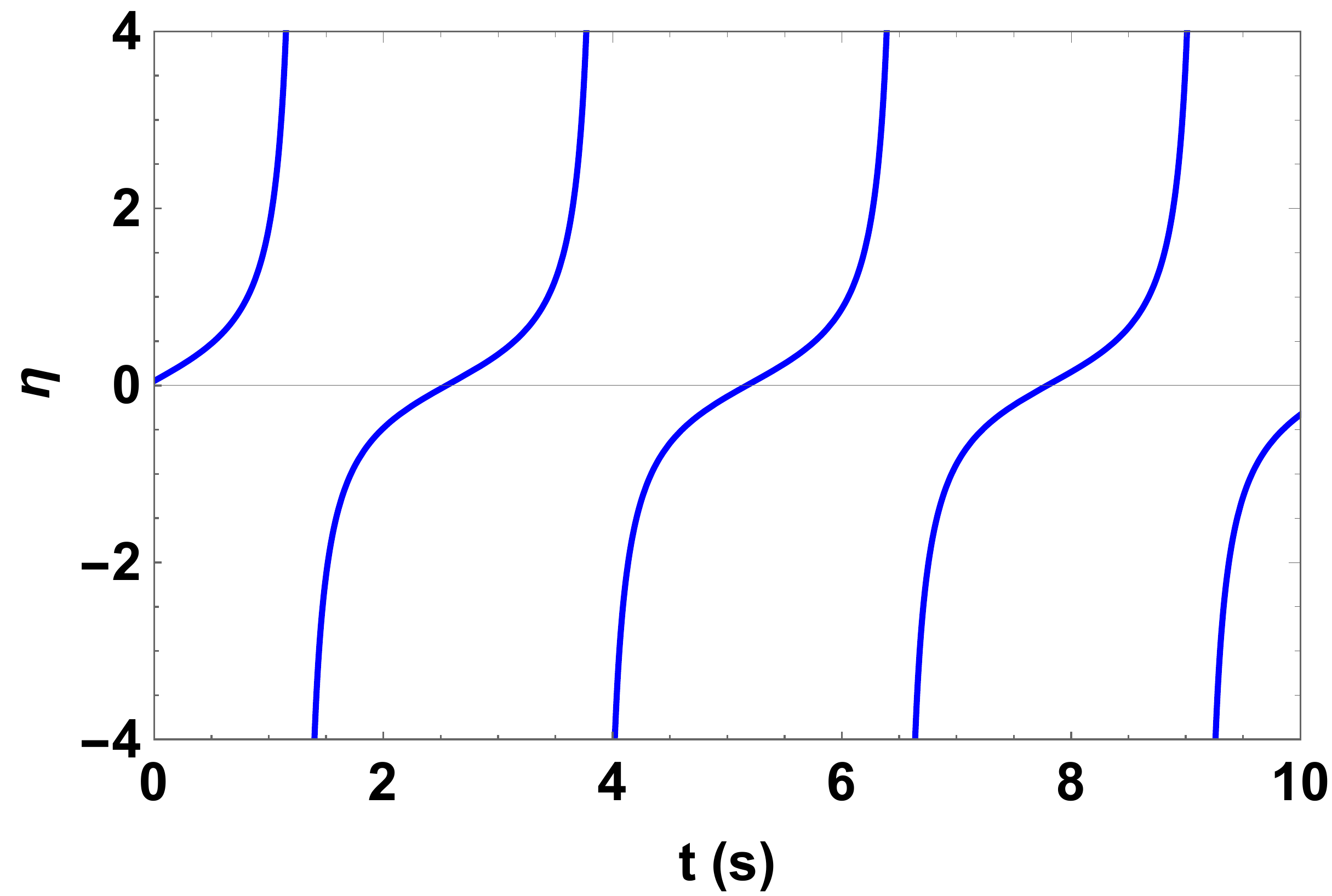}
	\caption{Plot of decoherence rate $\eta$ for RTN with $\gamma = 0.05$ and  $a=0.6$ (non-Markovian regime). The decoherence rate takes both positive and negative values.}
	\label{fig:gamma}
\end{figure}

Given a dynamical map $\mathcal{E}(t_2,t_0)$ linking a system's
density operator at times $t_0$ and $t_2 > t_0$, 
the intermediate map $\mathcal{E}^{\rm IM}(t_2,t_1)$
for some intermediate time $t_1$ such that  
$t_2 > t_1 > t_0$, is given by: 
\begin{equation}
	\mathcal{E}^{\rm IM}(t_2, t_1) = \mathcal{E}(t_2, t_0) \mathcal{E}^{-1}(t_1, t_0). 
	\label{eq:inter}
\end{equation}
provided that the inverse map
$\mathcal{E}^{-1}(t_1, t_0)$ exists. 
The Choi matrix for the intermediate map can be obtained as:
\begin{equation}
	\textrm{M}_{\rm Choi} = 
	(\mathcal{E}^{\rm IM}(t_2,t_1) \otimes \mathbb{I})\ket{\Phi^+}\bra{\Phi^+},
	\label{eq:mchoi}
\end{equation}
where $\ket{\Phi^+} \equiv \ket{00}+\ket{11}$.
Following  the method  presented in  \cite{RHP14}, we 
briefly derive the
intermediate dynamical  maps
of   the    RTN noisy   channel,  given   in   Eq.  \ref{noiseprop}. 
Now,   the   dephasing  master   equation   in   its  canonical   form 
\cite{andersson2007finding,NMDephasing1} is
\begin{equation}
	\dot{\rho} = \eta(-\rho + \sigma_3\rho \sigma_3),
	\label{eq:canondeph}
\end{equation}
where $\eta$  is the decoherence  rate.  It is  known that a  map is
CP-divisible if and only if the map has only positive decoherence rate
\cite{hall2014canonical}.
From the Kraus operator expressions in Eq. 
\ref{noiseprop},
we    obtain   $\rho    \rightarrow
\mathcal{E}(\rho)$. In particular, as the RTN noise is dephasing, we find:
\begin{equation}
	\rho_0 \equiv \begin{pmatrix}\rho_{00} & \rho_{01}\\
		\rho_{10} & \rho_{11} 
	\end{pmatrix}
	\rightarrow
	\rho(t) = \begin{pmatrix}\rho_{00} & \lambda\rho_{01}\\
		\lambda\rho_{10} & \rho_{11} 
	\end{pmatrix}.
\end{equation}
For this dephasing map, the decoherence rate is found by direct substitution in Eq. (\ref{eq:canondeph}) to be:
\begin{equation}
	\eta = -\frac{\dot{\lambda}}{2\lambda}.
	\label{eq:eta}
\end{equation}
For the channel under consideration
$\lambda = \Lambda(t)$, which, in view of Eq. (\ref{eq:lambda}), yield the decoherence rate for this channel, which is plotted in Fig. (\ref{fig:gamma}). We note that RTN shows negative decoherence rates for certain intervals.

\subsection{Distinguishability criterion}

We will consider the  composite dynamics comrising of the effect of RTN or NMAD on  an otherwise unitary  two-qubit dynamics governed by    Jayne-Cummings (JC)-like Hamiltonian.  

Trace   Distance  (TD)   \cite{laine2010measure}  is   a  measure   of distinguishability between two states,  defined as $D(\rho_1,\rho_2) = \frac{1}{2}\rm{Tr} \norm{\rho_1-\rho_2}$, where  $\norm{A}$ is the operator
norm  given by  $\sqrt{A^\dag{A}}$. It has been used as a measure of non-Markovianity to quantify the amount of backflow from the environment to the system. For  the noise  channel given  in
Eq. \ref{noiseprop}, and  initial states  $|\pm\rangle$, it  is
straightforward   to   compute  the   evolution   of   TD,  which   is
$D(\ket{+},\ket{-})  =  \Lambda(t)$  for  RTN. Thus,  we  find  that  the
distinguishability criterion, like the divisibility criterion, is able
to witness  the non-Markovianity  of RTN, because the process is P-indivisible.   A similar conclusion can be made for the NMAD channel, discussed above. We now consider the case where RTN or NMAD is applied to a qubit subject to another non-Markovian noise. Our study will then focus on how to distinguish the non-Markovian signatures of these two  composite processes through a simple frequency analysis.

\section{A simple model}

Consider a system of two qubits labeled as $A$ and $B$, where $A$ is the system and $B$ serves as environment. At time $t=0$, qubit $B$ is prepared in the state $\ket{\psi_B} = \frac{1}{\sqrt{2}} (\ket{0} + \ket{1})$, whilst qubit $A$ starts in  state $\ket{0}$ or $\ket{1}$. The two qubits interact via the  Jaynes-Cummings (JC) -like Hamiltonian \cite{scully-zubairy}
\begin{equation}
	H  = \omega \Big(| 01 \rangle \langle 10 | +  | 10 \rangle \langle 01 | \Big).
\end{equation}
The reduced dynamics of qubit $A$ can be described by  map $\mathcal{J}(t) : \rho_A (t) = \mathcal{J}[\rho_A (0)] = \sum_i J_i(t) \rho_A (0) J_i^\dagger(t)$, with the following Kraus operators
\begin{equation}\label{eq:mapJ}
	J_1 = \left(
	\begin{array}{cc}
		\frac{1}{\sqrt{2}} & 0 \\
		-\frac{i \sin (t \omega )}{\sqrt{2}} & \frac{\cos (t \omega )}{\sqrt{2}} \\
	\end{array}
	\right), \quad  J_2 = \left(
	\begin{array}{cc}
		\frac{\cos (t \omega )}{\sqrt{2}} & -\frac{i \sin (t \omega )}{\sqrt{2}} \\
		0 & \frac{1}{\sqrt{2}} \\
	\end{array}
	\right).
\end{equation} 
The time evolved reduced state of $A$ at time $t$, depending on whether it started in $\ket{0}$ or $\ket{1}$ at time $t=0$, is respectively given by:
%
\begin{align}\label{eq:rhoIB}
	\rho_{A}^{I} (t) &=  \left(
	\begin{array}{cc}
		\frac{1}{2} \cos ^2(t \omega )+\frac{1}{2} & \frac{1}{2} i \sin (t \omega ) \\
		-\frac{1}{2} i \sin (t \omega ) & \frac{1}{2} \sin ^2(t \omega ) \\
	\end{array}
	\right), \\
	\label{eq:rhoIIB}
	\rho_{A}^{II} (t) &=  \left(
	\begin{array}{cc}
		\frac{1}{2} \sin ^2(t \omega ) & -\frac{1}{2} i \sin (t \omega ) \\
		\frac{1}{2} i \sin (t \omega ) & \frac{1}{2} \cos ^2(t \omega )+\frac{1}{2} \\
	\end{array}
	\right).
\end{align}
We find the trace distance $\mathcal{D}(\rho^I_{A}(t), \rho^{II}_{A}(t)= \frac{1}{2} \sum_i |\lambda_i|$.       Here                     $\lambda_i$ are the eigenvalues of the matrix            $\rho^I_{A}(t) - \rho^{II}_{A}(t)$. We obtain
\begin{equation}\label{eq:TDnoiseless}
	\mathcal{D} =  \frac{\sqrt{7 + \cos (4  \omega t )}}{2 \sqrt{2}}
\end{equation}

The time evolution of qubit $A$ during the interval $(0,t)$ is described by a composite map $\mathcal{R}(t) \circ \mathcal{J}(t)$,   where $\mathcal{R}(t)$ denotes the RTN or NMAD channel, while $\mathcal{J}(t)$ indicates the dynamics generated by Eq. (\ref{eq:mapJ}). We have
\begin{align}\label{eq:TDsigma}
	\sigma^\alpha_A (t) &=( \mathcal{R}(t) \circ \mathcal{J}(t)) \rho_A(0),
\end{align} 
with $\alpha \in \{I, II\}$. 

The trace distance between these (Eq. (\ref{eq:TDsigma})) states,  for RTN, turns out to be 
\begin{equation}\label{eq:TDwithnoise}
	\mathcal{D} = \bigg| \frac{\sqrt{4 \Lambda ^2-4 \left(\Lambda ^2-1\right) \cos (2 t \omega )+\cos (4 t \omega )+3}}{2 \sqrt{2}} \bigg|.
\end{equation}
Note that the \textit{noiseless} case pertains to $\Lambda = 1$, and the above expression reduces to Eq. (\ref{eq:TDnoiseless}).

\begin{figure}[ht!]
	\centering
	\begin{tabular}{cc}
		\includegraphics[width=60mm]{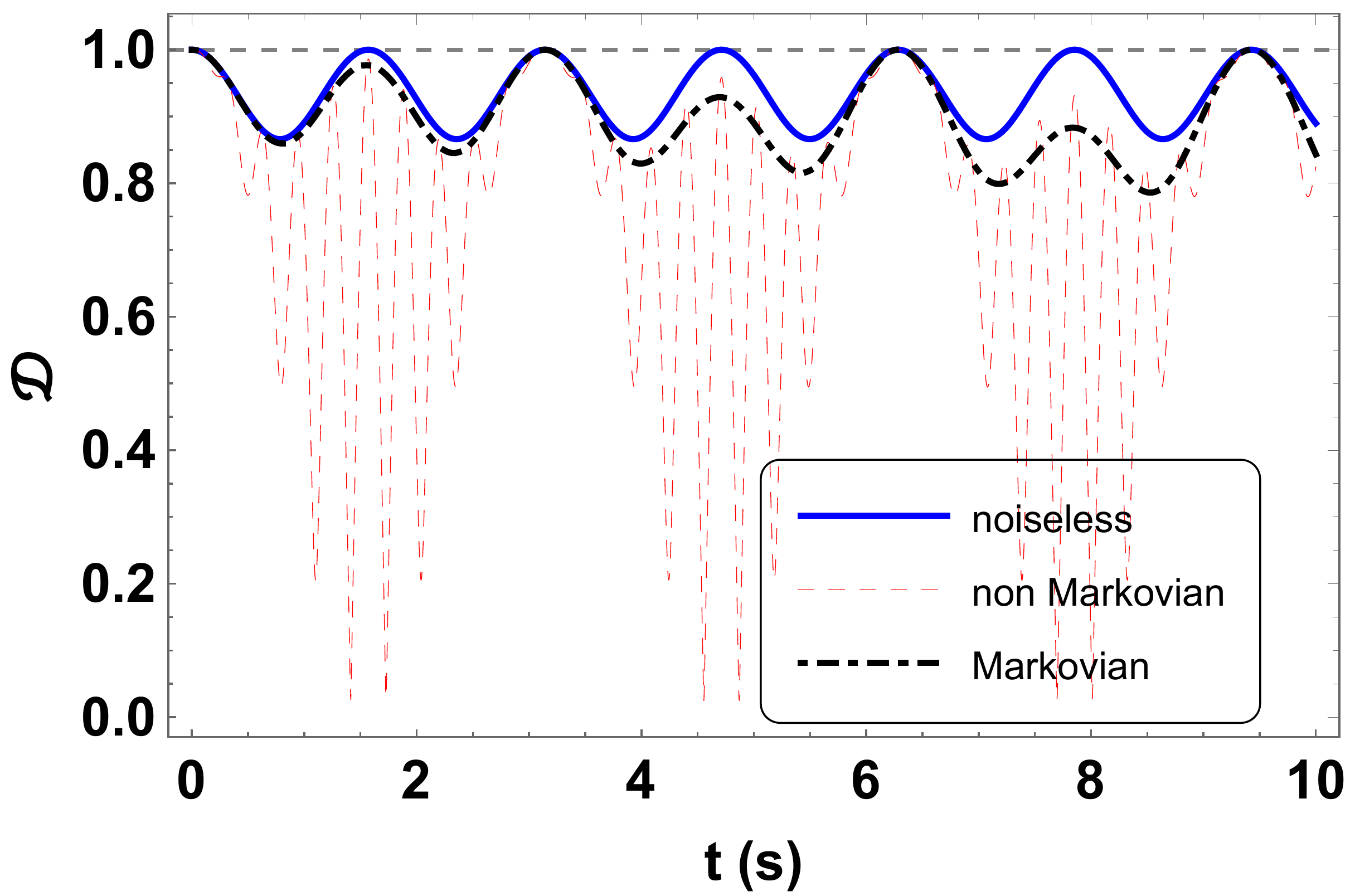} & \includegraphics[width=60mm]{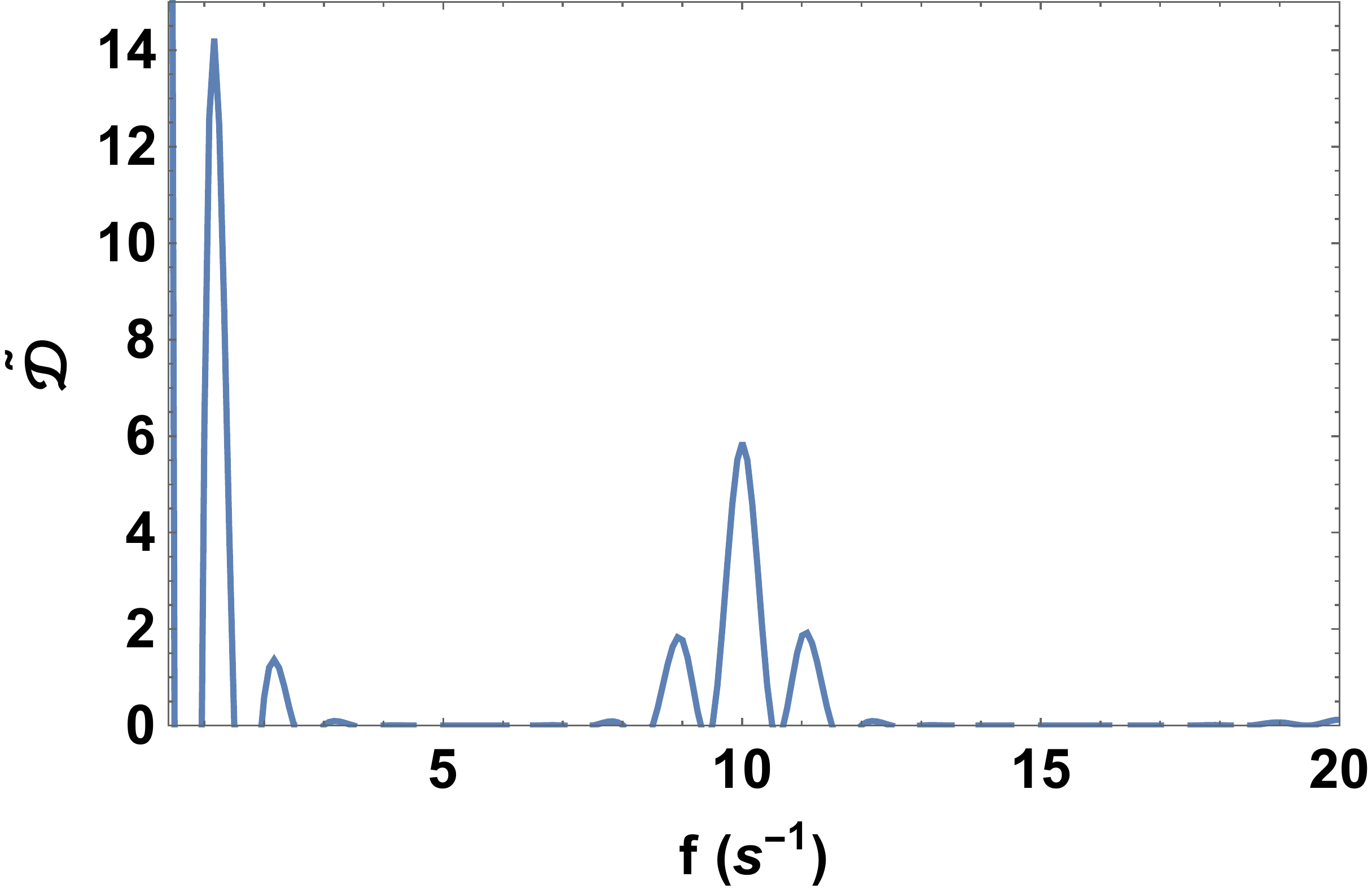} 
		\tabularnewline
		(a)  &  (b)  
	\end{tabular}
	\caption{ (a) Trace distance $\mathcal{D}$ as defined in Eq. (\ref{eq:TDwithnoise}) for noiseless case $\Lambda = 1$ (blue curve), for non-Markovian RTN case with $a = 5$, $\gamma = 0.009$, and for Markovian case $a = 0.2$, $\gamma = 5$.  In all the cases, the frequency $\omega$ which comes from the unitary evolution is set to one. (b) The Fourier transform $\tilde{\mathcal{D}}$ of trace distance function $\mathcal{D}$, defined by Eq. (\ref{eq:FT}), in non-Markovian regime. The dominating contribution to the frequency comes from  $ {\rm f} = 1$ from the noiseless dynamics and $ {\rm f} = 0.09$ and $10$ from RTN. }
	\label{fig:TDRTN}
\end{figure}

The form of the channel parameter  $\Lambda$ is given by Eq. (\ref{eq:lambda}) such that  $\frac{2a}{\gamma} > 1$ and $\frac{2a}{\gamma} < 1$    pertain  to the non-Markovian and Markovian cases, respectively. Figure (\ref{fig:TDRTN}) depicts the variation of the trace distance in different scenarios.\bigskip

Similar analysis performed  with the  NMAD channel,  Eqs. (\ref{eq:mapNMAD}), (\ref{eq:KrausA1A2}), leads to  following expression for  trace distance 
\begin{align}\label{eq:TDNMAD}
\mathcal{D} &=\frac{1}{2}\Bigg| \frac{1}{4} \big( 1 - G^2 - 2 \cos(2 \omega t) [-1 + G^2]\big) \nonumber \\&- \frac{1}{2} \sqrt{ \cos^4(\omega t) [-1 + G^2]^2 + \frac{G}{2} \big[  (7 + \cos(4\omega t)) G + 8[-1+G^2] \sin^2(\omega t)\big]} \Bigg| \nonumber \\&+ \frac{1}{2} \Bigg| \frac{1}{4} \big( 1 - G^2 - 2 \cos(2 \omega t) [-1 + G^2]\big) \nonumber \\&+ \frac{1}{2} \sqrt{ \cos^4(\omega t) [-1 + G^2]^2 + \frac{G}{2} \big[  (7 + \cos(4\omega t)) G + 8[-1+G^2] \sin^2(\omega t)\big]} \Bigg|.
\end{align}
For $G = 1$, the Kraus operators, given in Eq. (\ref{eq:KrausA1A2}), become $A_1 = I$, $A_2 = 0$, and we recover the noiseless  Eq. (\ref{eq:TDnoiseless}).

\section{Disambiguation of multiple non-Markovian effects}

\begin{figure}[ht!]
	\centering
	\begin{tabular}{cc}
		\includegraphics[width=60mm]{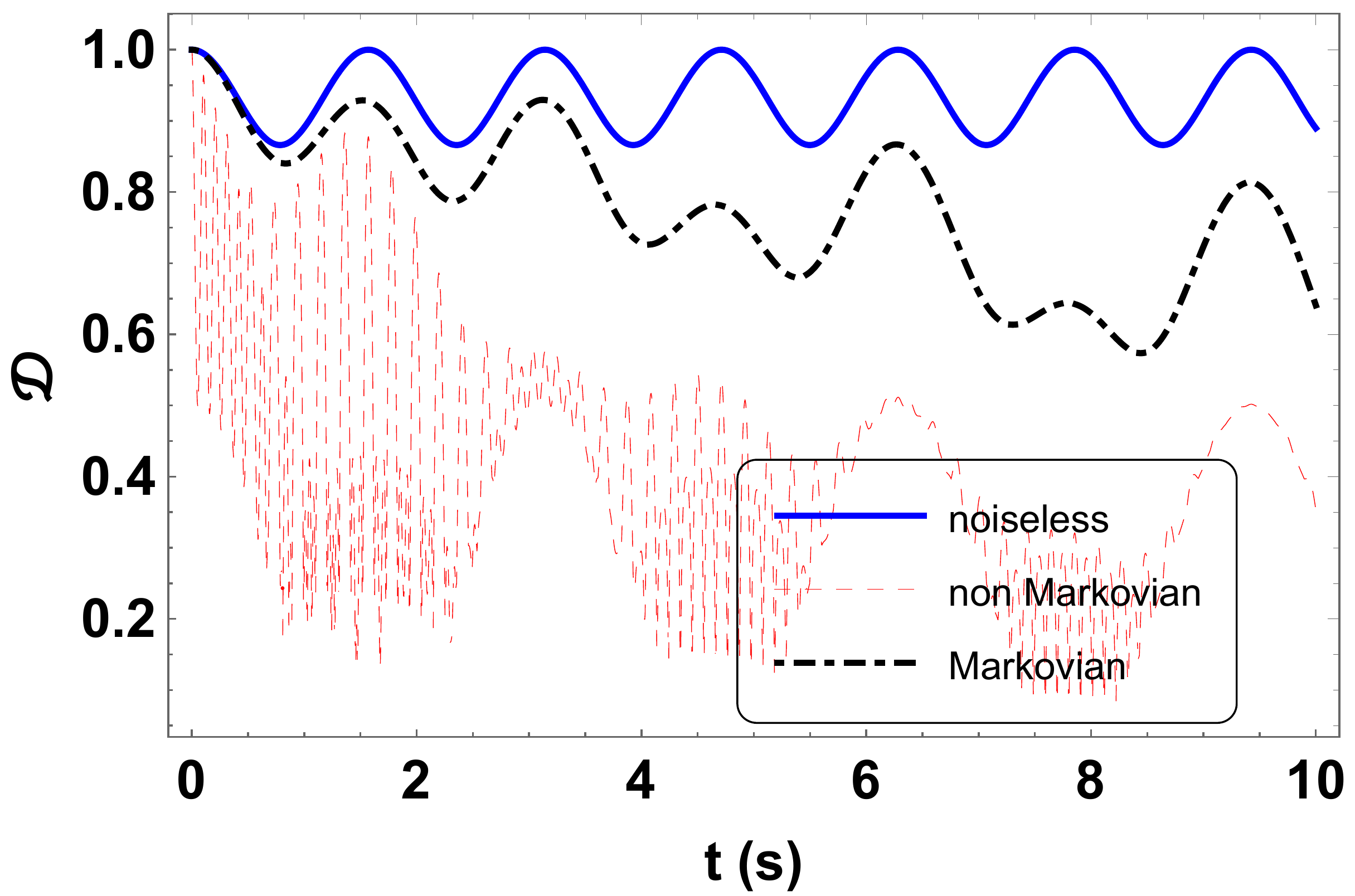} & \includegraphics[width=60mm]{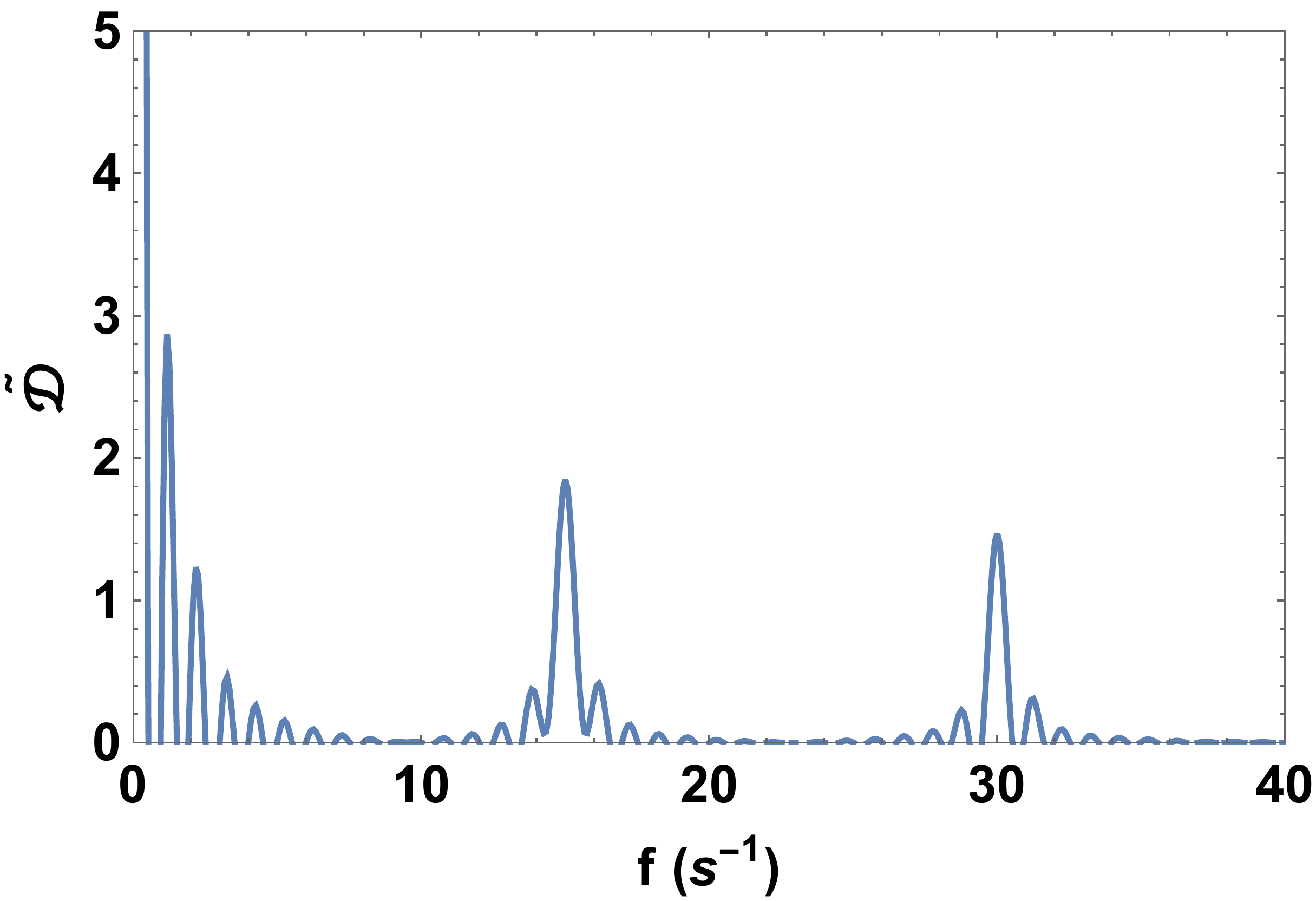} 
		\tabularnewline
		(a)  &  (b)  
	\end{tabular}
	\caption{ (a) Trace distance $\mathcal{D}$ as defined in Eq. (\ref{eq:TDNMAD}) for noiseless case $G(t) = 1$ (blue curve), for Markovian AD case with $\lambda= 10, \gamma_M = 0.1$, and for non-Markovian case $\lambda = 0.6, \gamma_M = 100$.  In all the cases, the frequency $\omega$ which comes from the unitary evolution is set to one. (b) The Fourier transform $\tilde{\mathcal{D}}$ of trace distance function $\mathcal{D}$ defined by Eq. (\ref{eq:TDNMAD}). The  frequencies   $ {\rm f} = 0.3$ and $30$ correspond to  NMAD, and  ${\rm f} = 1$  to the noiseless dynamics. }
	\label{fig:TDNMAD}
\end{figure}

Here,  we concern ourselves with   the  situation   when non-Markovianity can have multiple sources of noise. In \cite{pradeep1}, one is an internal source (the coin degree of freedom) and the other an 	external  source   (external  environment). For the system of  two qubits driven by a JC-type of Hamiltonian, considered above, the oscillatory behavior of the trace distance with respect to time, as given in Eq. (\ref{eq:TDnoiseless}), reflects the non-Markovian character of the the dynamics. On top of this, we subject the system to the  RTN or NMAD noise. Consequently, the trace distance, given by Eq. (\ref{eq:TDwithnoise}), now acquires a dependence on two frequency components, one from the subsystem dynamics and the other from the environmental noise, as depicted in Figs. \ref{fig:TDRTN} (a) and \ref{fig:TDNMAD} (a), for the RTN and NMAD cases, respectively.

The different source contributions to non-Markovian dynamics can be demarcated by using  simple Discrete Fourier Transform (DFT). The trace distance, Eq. (\ref{eq:TDwithnoise}) and (\ref{eq:TDNMAD}), apart from the frequency $\omega$, which comes from dynamics governed by  JC-like two qubit Hamiltonian,  also depends on the memory kernels $\Lambda$ of RTN and $G$ of the NMAD noise, respectively. The form of $\Lambda$, as given in Eq. (\ref{eq:lambda}), reveals the dependence on two frequencies $\gamma$ and $\gamma \sqrt{(2 a/\gamma)^2 - 1}$.   Figures \ref{fig:TDRTN} (b), \ref{fig:TDNMAD} (b), depict the DFT of the trace distance defined in Eq. (\ref{eq:TDwithnoise}) and (\ref{eq:TDNMAD}), respectively,  which brings out a reasonably clear separation of the different contributing factors. The corresponding power spectrum is given by $|\tilde{\mathcal{D}}(f)|^2$, where $\tilde{\mathcal{D}}$ is the Fourier transform of the $\mathcal{D}(t)$, given by
\begin{eqnarray}
	\tilde{\mathcal{D}}(f) = \int \mathcal{D}(t) e^{-2\pi i ft}dt,
	\label{eq:FT}
\end{eqnarray}
implemented numerically. As the two noise contributions ( JC-like dynamics and RTN or NMAD) occur at different frequencies, the TD spectral power at those frequencies can be directly read off as representing the relative strength of their contribution to the non-Markovianity. The power spectral plot in Figure \ref{fig:TDRTN} (b) shows that the dominant contribution is seen to to be at the frequency of the RTN with $ \gamma = 0.009$, $\gamma \sqrt{(2 a/\gamma)^2 - 1} = 10$, with a lower contribution due to JC-like dynamics with $\omega = 1$ (again, identified by the associated frequency).  Similarly, when the system is subjected to NMAD channel, one finds two dominating contributions to the frequency spectrum, at  $(\lambda - i \delta)/2 = 0.3$ and $\Omega/2 = \frac{1}{2}\sqrt{\lambda^2 - 2 i \delta \lambda - 4w^2 } =  30$ coming from  NMAD dynamics, and around $\omega = 1$ due to noiseless dynamics. Further, the frequency ${\rm f =15}$ is also seen to contribute actively, which may be a consequence of the complicated structure of the memory kernel $G(t)$, Eq. (\ref{eq:Gt}).
\bigskip

\begin{figure}[ht!]
	\centering
	\begin{tabular}{cc}
		\includegraphics[width=60mm]{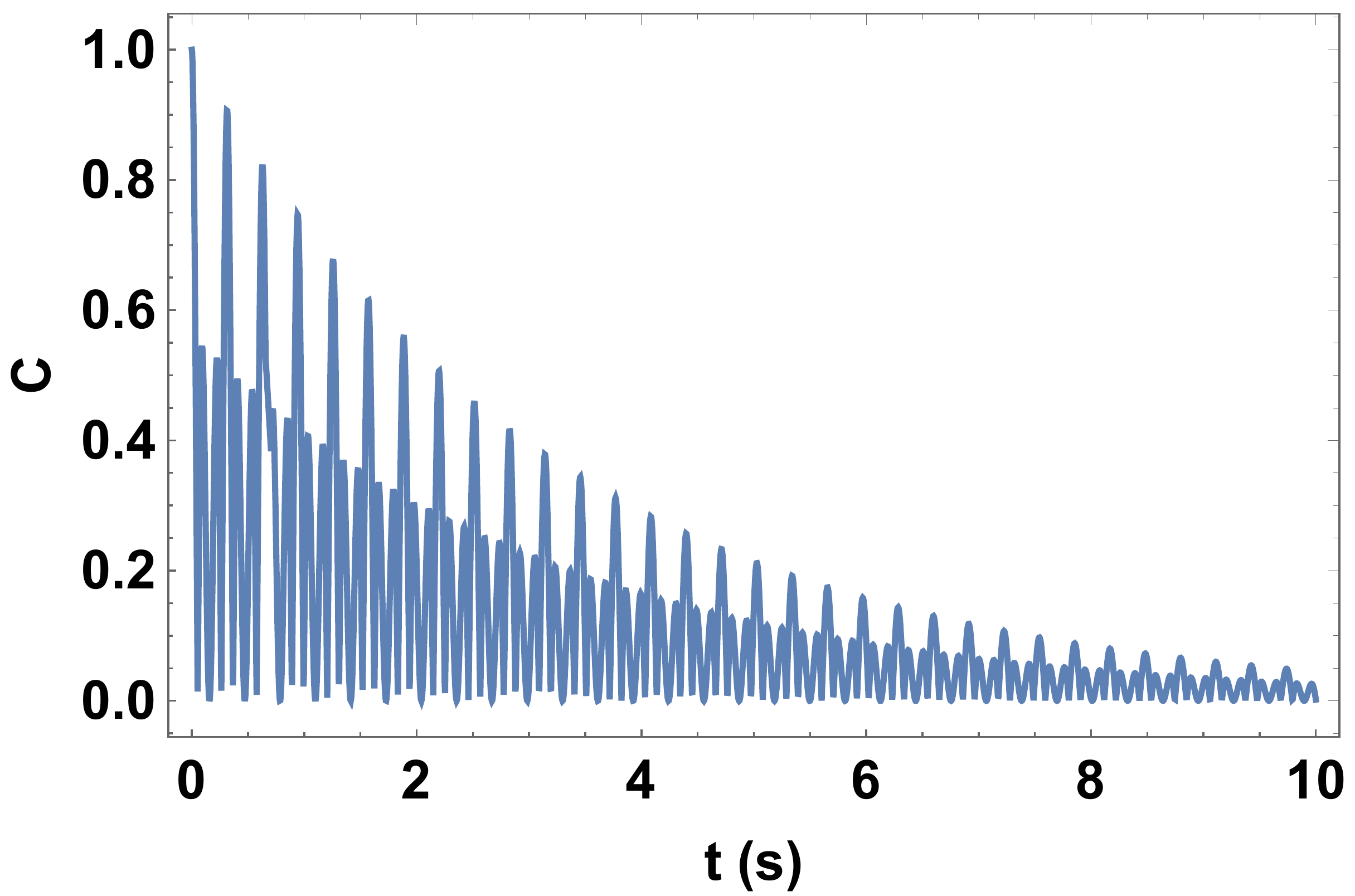} & \includegraphics[width=60mm]{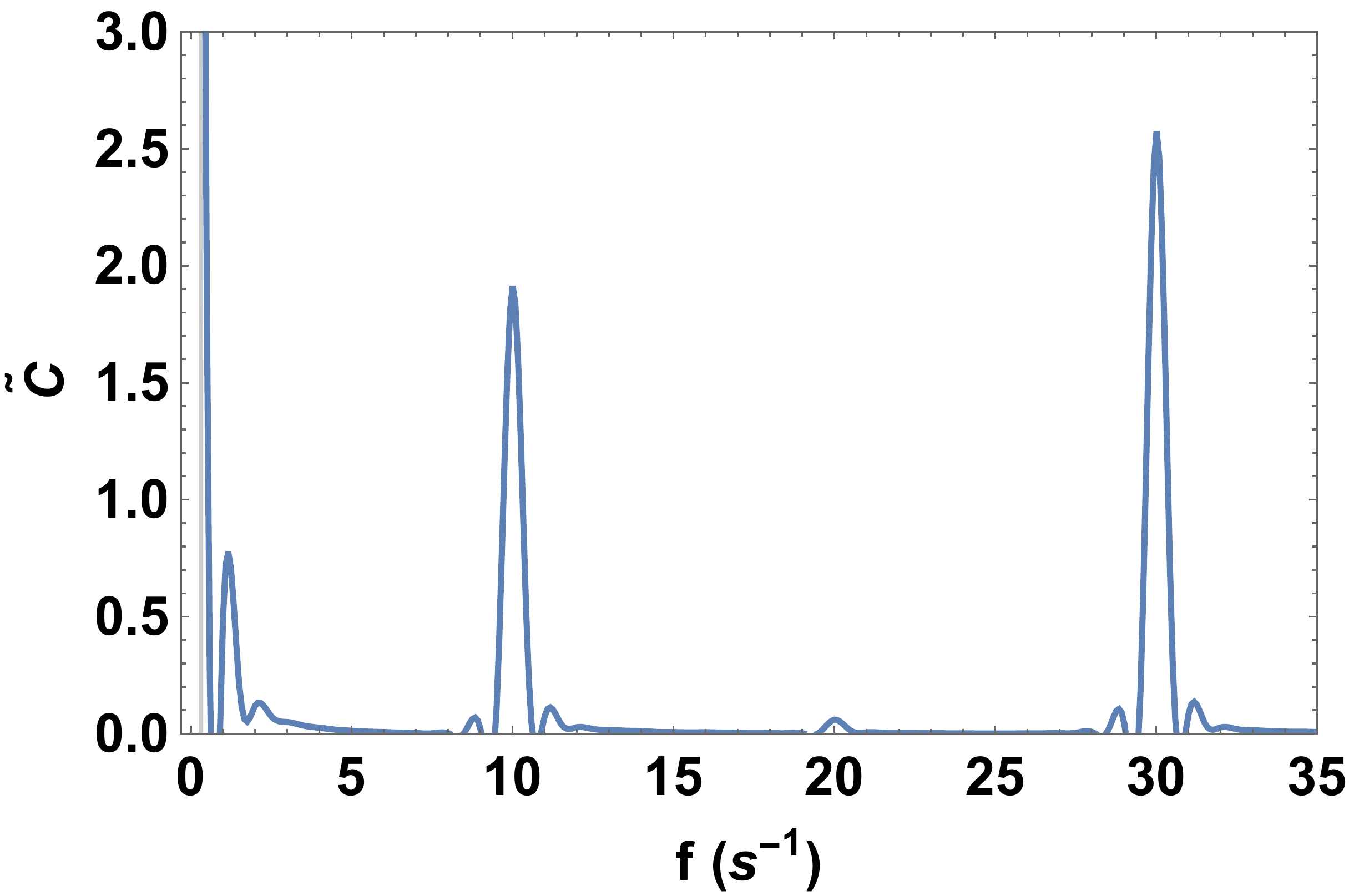} 
		\tabularnewline
		(a)  &  (b)  
	\end{tabular}
	\caption{ Depicting coherence $C$ as defined in Eq. (\ref{eq:Coherence}) (a) and its Fourier transform $\tilde{C}$ (b),  when a qubit is subjected to  dynamics generated by composition of RTN and NMAD channels both operating in non-Markovian regime. The various parameters used are the same as in Figs. (\ref{fig:TDRTN}) and (\ref{fig:TDNMAD}).  The dominating contribution to the frequency due to RTN is at   $ {\rm f} = 0.01$, and  $10$, and from NMAD at  $ {\rm f} = 0.3$ and $30$. }
	\label{fig:coherence}
\end{figure}

 \textit{A more complicated scenario:} We now consider a more complex situation wherein the  qubit state $\rho$ is acted upon by the composition of RTN and NMAD channels as 
\begin{equation}\label{eq:compositeMap}
\rho^\prime = \sum_{\mu,\nu = 1,2} K_\mu A_\nu \rho A_\nu^\dagger K_\mu^\dagger.
\end{equation}
Here, $K_\mu$ and $A_\nu$ are defined in Eqs. (\ref{noiseprop}), and (\ref{eq:KrausA1A2}), respectively.  Trace distance is a useful witness for non-Markovian dynamics, as seen above in case of RTN and NMAD. Here, we analyze  the effect of memory due to the composite map on the coherence of a qubit state. We use the  $l_1$ norm definition of coherence given by the sum of absolute values of the off-diagonal elements of density matrix, i.e., $C = \sum_{i \ne j} |\rho_{ij}|$ \cite{SBcoh,javid,KDixit}.  With  initial state $\rho = \ket{+}\bra{+}$ subjected to the composite map, Eq. (\ref{eq:compositeMap}),  the coherence of output state turns out to be 
\begin{equation}\label{eq:Coherence}
C = \Lambda (t) G(t),
\end{equation}
where $\Lambda(t)$ and $G(t)$ are defined in Eqs. (\ref{eq:lambda}), and (\ref*{eq:Gt}), respectively. Thus, coherence is given by the product of two memory kernels, capturing the non-Markovian features of both the channels.  In Fig. (\ref{fig:coherence}) is depicted the coherence and its Fourier transform. The frequencies ${\rm f}= 0.1$, and $10$ come from RTN and ${\rm f} = 0.3$ amd $ 30$ from NMAD dynamics. A clear disambiguation of the effect of  two channels is observed. Further, with fundamental frequency of $f_0=15$, one observes higher harmonics at $n f_0$, ($n=2,3,\dots$). This  invites further investigation in more complicated scenarios.

\section{Conclusion}  

Quantum non-Markovianity is a fundamental memory effect in open system
dynamics. A simple manifestation thereof is a  departure from  the  monotonic fall  of distinguishability
between states, or  of the CP-divisibility of  the system's evolution.
In  complex  systems, subsystem  dynamics  can  compound this  effect by having noisy contributions due to different factors.
Here, we studied the non-Markovian effects on the evolution of a qubit subjected to two types of noise: one due to a JC-type interaction with a second qubit, and another due to RTN (NMAD) channel.

Both noisy channels are P-indivisible, and thus the distinguishability criterion is sufficient to study their effect. We derived the trace distance dependence of the system's evolution on characteristic frequencies for the  JC-type dynamics induced noise and RTN (NMAD) channel. Their corresponding frequencies can be demarcated using simple discrete Fourier transform,  distinguishing  non-Markovian contributions to the dynamics coming from the  JC-like dynamics source  and the external (environment) source.  The technique is also seen to work in the more complicated case of a qubit subjected to a concatenation of  RTN and NMAD noise channels, both realistic noises arising via interactions with proper baths. In the later case, a fundamental frequency and it harmonics are found to occur in the power spectrum which invites for further investigation in more complicated scenarios. They may present challenges that would make the unraveling of the time-scales harder. It would be interesting to extend the present work to incorporate such cases.


\end{document}